# Testing Microfluidic Fully Programmable Valve Arrays (FPVAs)

Chunfeng Liu[1,5], Bing Li[1], Bhargab B. Bhattacharya[2], Krishnendu Chakrabarty[3,5], Tsung-Yi Ho[4,5], Ulf Schlichtmann[1]
[1]Institute for Electronic Design Automation, Technical University of Munich, Germany   [2]Indian Statistical Institute, Kolkata, India
[3]Department of ECE, Duke University, Durham, NC, USA   [4]National Tsing Hua University, Hsinchu, Taiwan
[5]Institute for Advanced Study, Technical University of Munich, Germany
{chunfeng.liu, b.li, ulf.schlichtmann}@tum.de, bhargab@isical.ac.in, krish@ee.duke.edu, tyho@cs.nthu.edu.tw

*Abstract*—Fully Programmable Valve Array (FPVA) has emerged as a new architecture for the next-generation flow-based microfluidic biochips. This 2D-array consists of regularly-arranged valves, which can be dynamically configured by users to realize microfluidic devices of different shapes and sizes as well as interconnections. Additionally, the regularity of the underlying structure renders FPVAs easier to integrate on a tiny chip. However, these arrays may suffer from various manufacturing defects such as blockage and leakage in control and flow channels. Unfortunately, no efficient method is yet known for testing such a general-purpose architecture. In this paper, we present a novel formulation using the concept of flow paths and cut-sets, and describe an ILP-based hierarchical strategy for generating compact test sets that can detect multiple faults in FPVAs. Simulation results demonstrate the efficacy of the proposed method in detecting manufacturing faults with only a small number of test vectors.

## I. INTRODUCTION

Microfluidic biochips have revolutionized the traditional slow and error-prone biochemical experiment flow by manipulating nanoliter volumes of fluids precisely [1], [2], [3]. With this miniaturization, bioassays can be scaled down and genomic bioassay protocols, such as nucleic-acid isolation, DNA purification, and DNA sequencing, have been successfully demonstrated with these chips. In addition, this technology has also attracted a lot of commercial attention, e.g., from Illumina [4], a market leader in DNA sequencing.

Microfluidic biochips based on continuous flow use valves to control the movement of samples and reagents. The structure of a valve is shown in Fig. 1(a). In such a structure, a flow channel is constructed on a substrate for transportation of fluids. Above the flow channel, a control channel is constructed and connected to an air pressure source. Since both channels are built from elastic materials, air pressure applied in the control channel squeezes the flow channel tightly, so that the movement of the fluid is blocked. Conversely, if the pressure in the control channel is released, the fluid can resume its movement to the target destination. Consequently, a valve is formed at the intersection of the two channels.

Valves can also be used to build complex devices. For example, the structure of a mixer is shown in Fig. 1(b). When the three valves at the top of the mixer are actuated alternately by applying and releasing air pressure in the control channels, a circular flow around the device can be formed to mix different samples and reagents. After an operation is completed, the intermediate result can be transported to other devices or stored temporarily in a dedicated storage unit. Fig. 1(c) shows a detailed schematic of a mixer connected to a storage unit with eight cells. These neighboring storage cells can be constructed using normal flow channels and multiplexed control valves at both ends.

Recent advances in manufacturing technologies have enabled valve density to reach 1 million per $cm^2$ [6], and consequently,

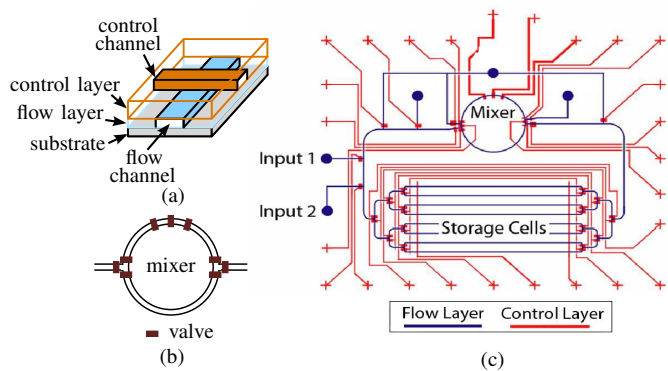

Fig. 1. Components and structure of flow-based biochips. (a) Valve structure. (b) Mixer. (c) Biochip with eight storage cells [5].

fully programmable valve arrays (FPVAs) have emerged for more flexible and highly reconfigurable flow-based biochips [1], [7]. Fig. 2(a) shows a part of the large valve array demonstrated in [7]. In this architecture, valves (solid blocks) are arranged in a regular manner along horizontal and vertical flow channels (light color). These valves are controlled by air pressure sources through the control channels (narrow channels). By opening two valves and closing the other two at a crosspoint of flow channels, the fluid sample stored there can be moved in the intended direction by forming temporary transportation channels.

Besides transportation channels, complex devices such as mixers can be constructed on the valve array by taking advantage of the flexibility and reconfigurability of such chips [8]. For example, a 4×2 mixer and a 2×4 mixer can be constructed as in Fig. 2(b) and 2(c), respectively. In such a dynamic mixer, the eight valves along the enclosed channel function as pump valves, which switch in a given pattern to drive the fluid samples and reagents inside the channel for mixing. Compared with the traditional mixer shown in Fig. 1(b), these dynamic mixers have a different shape and more pump valves, eight in each case, to form a strong circular mixing flow. The two mixers in Fig. 2(b) and 2(c) can share the same part of the chip area as shown in Fig. 2(d), provided that they are not used at the same time. Consequently, the same area of a valve array can execute various functions such as mixing and flow transportation, as well as detection if the corresponding sensors are included in this area.

It is convenient to fabricate FPVAs as large-scale integrated devices, because a regular structure is easy to design and manufacture compared with the traditional irregular fluidic architecture, similar to the case of DRAM-arrays in the semiconductor industry. In addition, dynamic reconfigurability enables the valve array to execute nearly any application. This flexibility allows chip vendors to focus on improving the integration scale without worrying about the applications. On the other hand, customers who use such chips also have the flexibility to perform different applications, a notable advantage specially for small healthcare centers or research laboratories that usually cannot afford expensive apparatus.

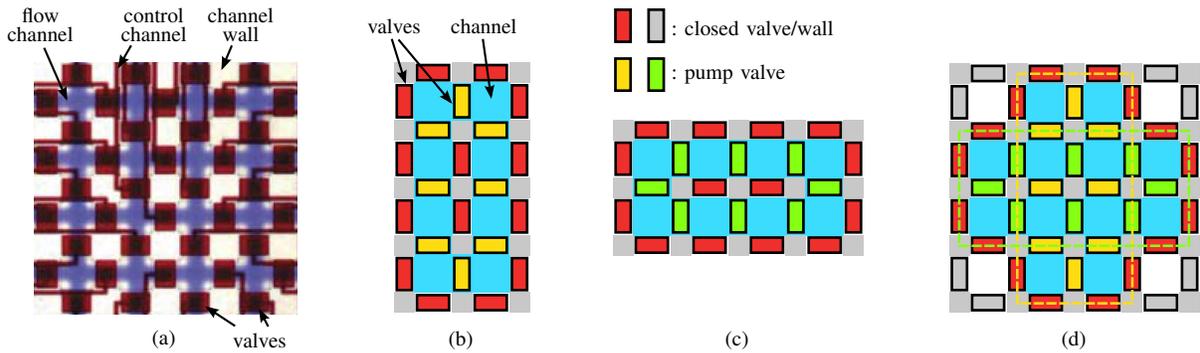

Fig. 2. Fully programmable valve array (FPVA). (a) Architecture [7]. (b)/(c) A 4×2/2×4 dynamic mixer. (d) Dynamic mixers of different orientations sharing the same area.

In order to facilitate industry adoption of FPVAs, efficient defect-screening techniques must be developed. In this paper, we focus on testing of FPVAs to identify chips with manufacturing defects. Our contributions are as follows:

- We propose the first systematic formulation and test strategy for detecting manufacturing defects efficiently in FPVAs;
- Our method is based on the construction of flow paths and cut-sets in the array, followed by a hierarchical ILP-based solution, which yields a very compact test set;
- The proposed method can guarantee the detection of up to two faults in the chip, while more than two faults can also be detected in nearly all cases;
- The method is general. It works both for a full array and an incomplete one with fluidic-seas (channels) or obstacles;
- The proposed test flow is compatible with test flows for traditional flow-based biochips, so that no additional cost is incurred for testing FPVAs.

The rest of this paper is organized as follows. In Section II, we review prior work on testing of traditional flow-based biochips and formulate the test problem for FPVAs. In Section III, we describe the general test strategy and explain how this strategy is implemented. Simulation results are reported in Section IV. Conclusions are stated in Section V.

## II. FAULT MODEL AND PROBLEM FORMULATION

During the manufacturing of FPVAs, various defects may occur, as illustrated in Fig. 3. These defects have been analyzed in detail and the corresponding fault models have been defined in [9]. Based on how these defects affect the behavior of a valve or a channel, faults at the component level can be defined as follows:

- *A break in a flow channel*: Fluid cannot pass through a channel. This is equivalent to the fault that the valve at the entrance of the channel cannot be opened.
- *A break in a control channel*: Air pressure cannot reach a valve to close it.
- *Leaking flow channel*: Fluid in a channel leaks to neighboring channel. In FPVA test, this fault is similar to the fault that a valve cannot be closed, because there is always a valve between two channels.
- *Leaking control channel*: Two valves close simultaneously due to the shared pressure in the control layer.

In the following discussion, if a valve cannot be opened, we refer to the scenario as a *stuck-at-0 fault*. Similarly, if a valve cannot be closed, we refer to the scenario as a *stuck-at-1 fault*.

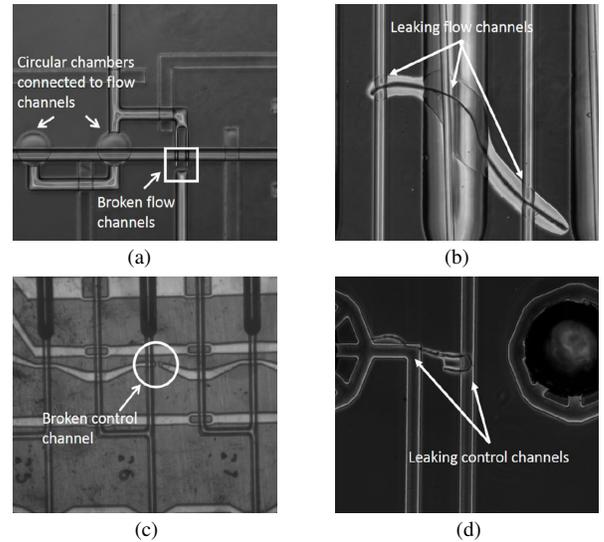

Fig. 3. Manufacturing defects in flow-based biochips [9]: (a) Broken flow channel; (b) Leaking flow channel; (c) Broken control channel; (d) Leaking control channel.

To detect faults in a flow-based biochip, the method in [9] connects air pressure sources to the input ports of the biochip to generate a pressure in the flow layer. By switching some valves open or closed with different control input vectors, air pressure patterns at the output ports can be used to deduce whether there is a fault in the chip. For example, in Fig. 4, a pressure can be detected at the output $o_2$ if the valves $a$, $g$, $h$, $i$, $k$ are open, while the other valves are closed. However, if there is a valve on this path that cannot be opened, there is no pressure at $o_2$, indicating the existence of a stuck-at-0 fault. On the other hand, if a valve on this path, e.g., $g$, is closed, while the other valves are open, there is no path from the pressure source to the output ports. If a pressure can still be detected at the output port, at least one stuck-at-1 fault exists.

In the above analysis, it can be seen that test results at the output ports carry fault information of internal valves, similar to the testing of integrated circuits. Therefore, the method in [9] represents the relationship between paths and valves using a digital circuit model and generates ATPG vectors to identify faults. This method offers the advantage of adopting a mature test flow by mapping the test problem to an ATPG problem. It, however, faces the challenge of generating a very large circuit model or many circuit models for testing an FPVA, because valves in an FPVA can be reconfigured to form a huge number of dynamic chip architectures, all of which should be covered by the ATPG-based method.

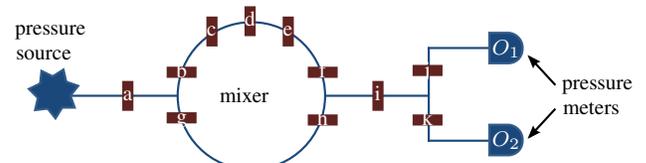

Fig. 4. Example of a biochip under test.

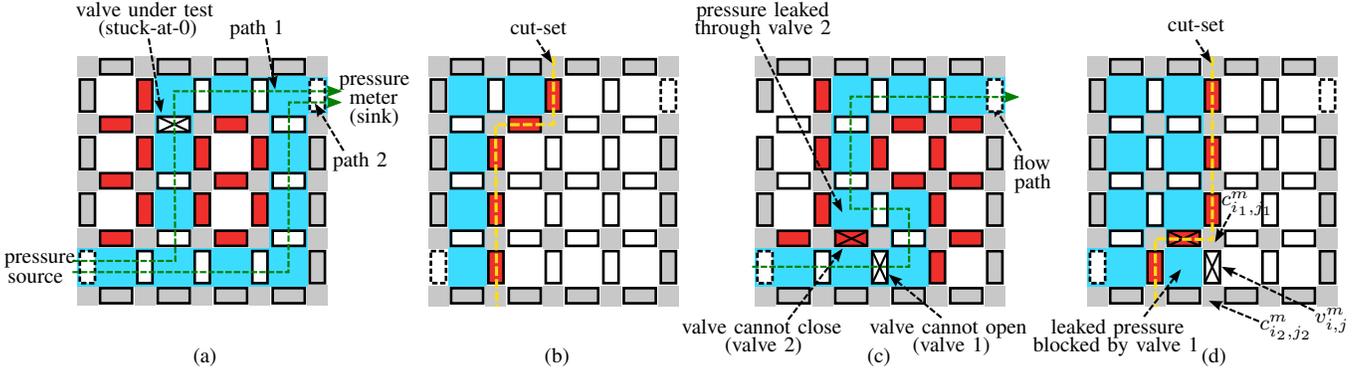

Fig. 5. Flow paths and cut-sets. Valves at the external boundary of the chip are always closed. (a) Flow paths masking a stuck-at-0 fault. (b) Cut-set. (c) Flow path with two-fault masking. (d) Cut-set with two-fault masking.

In this paper, we propose a new test framework that can detect faults in a manufactured chip reliably with only a small set of test vectors. This problem can be formulated as follows:

*Inputs*: An FPVA architecture; the locations of valves that are not built on flow channels (conceptually always open), and the locations of obstacles (conceptually always closed); the locations of the air pressure sources and the pressure meters.

*Outputs*: A set of test vectors, where each vector defines the open/closed states of all valves when test pressure is applied at the sources and checked at the output ports by the pressure meters.

*Objectives*: The number of test vectors to detect faults in an FPVA should be as small as possible to reduce test cost. The number of undetected faults should be as small as possible.

### III. FPVA TEST WITH FLOW PATHS AND CUT-SETS

In this section we explain the strategy to test an FPVA and the implementation of the test method. Although omitted from this paper due to the space limit, leakage at the control layer can also be detected by adapting the valve coverage problem described in this section. For convenience, we refer to a pressure source simply as a source port, and to a pressure meter port as a sink port in the following discussion.

#### A. Test strategy for an FPVA

To identify whether a fault exists in a chip, the test vectors should ensure that an error is observable when valves are switched. For example, if no test vector opens a valve during the test process, the only fault that can be observed at this valve is the leakage fault (stuck-at-1 fault) and the fault that the valve cannot be opened (stuck-at-0 fault) is not tested. Therefore, the test vectors should switch a valve open at least once and closed at least once during test application to detect stuck-at-0 and stuck-at-1 faults at this valve. For each case, the effect of the correct behavior of the valve should be observable at sink ports. For example, if a valve is switched open, there should be at least one path from the source through this valve to the sink to detect the pressure being transmitted through this valve.

When test vectors are applied, some faults might mask each other. For example, if another path that circumvents the valve under test connects the pressure source and the sink port, the potential stuck-at-0 fault (always closed) at the valve under test may be masked and thus cannot be observed. This situation is illustrated in Fig. 5(a). In this example, the two paths are created on the chip at the same time, and thus the stuck-at-0 fault at the valve under test cannot be detected because there is still a pressure at the sink port due to the second path. To avoid this path interference problem, we only construct simple paths without loops or branches. These paths are called *flow paths* henceforth.

Similar to constructing flow paths to detect stuck-at-0 faults, we construct cut-sets to detect stuck-at-1 faults. A *cut-set* is formed by a set of valves that separate the source ports and the sink ports completely. In test application, if all the valves in a cut-set are closed and a pressure is still detected by a pressure meter, a stuck-at-1 fault must exist. An example of such a cut-set is illustrated in Fig. 5(b), which disconnects any path between the source and the sink.

In a scenario with multiple faults, the flow-path and cut-set vectors discussed above, however, cannot guarantee that a fault can always be detected. Assume that there are two faulty valves, one of which cannot be opened (valve 1, stuck-at-0) and the other cannot be closed (valve 2, stuck-at-1). Also assume that the flow path used to test valve 1 is constructed as shown in Fig. 5(c), and the cut-set used to test valve 2 is constructed as shown in Fig. 5(d). In Fig. 5(c), the pressure leakage through valve 2 masks the stuck-at-0 fault at valve 1. In Fig. 5(d), the pressure leakage through valve 2 is blocked by valve 1. In both cases, the results at the pressure meter are still correct, so that these two faults cannot be detected. Consequently, mutual masking patterns of this type should be excluded from the generated test vectors.

#### B. Generating flow-path test vectors

In the proposed method, we generate the flow paths using an Integer Linear Programming (ILP) model. The scalability of this model is improved using a hierarchical approach.

##### 1) Constructing flow paths

A fluid cell is defined as the channel area surrounded by four valves, as shown in Fig. 6(a). Air pressure through a cell must pass through two of the valves surrounding this cell. Since the air pressure can reach the cell in any direction, in total there are 12 possible directions for a path passing through this cell as shown in Fig. 6(a). Instead of modeling theses directions directly, we model how the path passes through the surrounding valves. Suppose all valves can be covered by no more than $n_p$ flow paths in the test set, where $n_p$ is a given constant. For the cell at the location of the $i$th row and the $j$th column of the valve array, we assign a 0-1 variable $c_{i,j}^m$ to represent whether the $m$th path passes through the cell. If the $m$th path passes through the cell, $c_{i,j}^m=1$; otherwise $c_{i,j}^m=0$. For the valves at the left, right, upper and lower sides of the cell at the location $(i,j)$, we assign 0-1 variables $v_{i,j-1}^m$, $v_{i,j+1}^m$, $v_{i+1,j}^m$ and, $v_{i-1,j}^m$, respectively. If the $m$th path passes through a valve, the corresponding variable is set to 1; otherwise, it is set to 0.

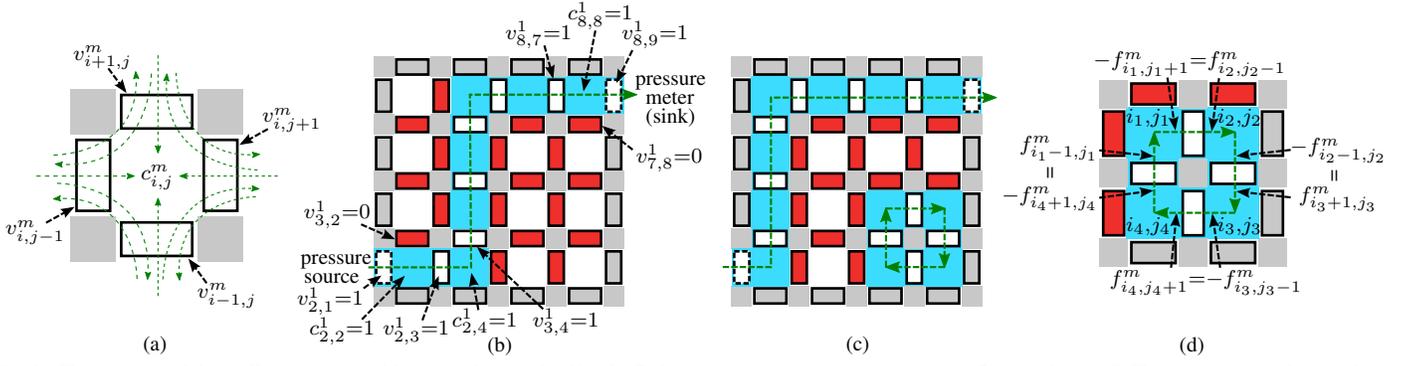

Fig. 6. Flow path model. (a) Constraint variables for valves and cells. (b) Path construction using constraints. (c) Disjoint loop. (d) Flow constraints along a disjoint loop.

If the $m$th path passes through the cell at the location $(i,j)$, this path should pass through exactly two valves that surround the cell; otherwise, no valve surrounding the cell should be passed. Consequently, the relation between the cell and the valves surrounding it can be established as

$$v_{i,j-1}^m+v_{i,j+1}^m+v_{i+1,j}^m+v_{i-1,j}^m=2c_{i,j}^m, \forall\ i=1,2,...,n_r, \quad j=1,2,...,n_c,\ m=1,2,...,n_p \quad (1)$$

where $n_r$ and $n_c$ are numbers of rows and columns of the valve array, respectively. Constraint (1) constructs the $m$th path by the chaining effect of the variables $v_{i,j}^m$ as demonstrated in Fig. 6(b).

To guarantee that a valve is covered at least once by the flow paths, one of the constraint variables $v_{i,j}^m$ for the valve indexed by $(i,j)$ on the $m$ paths must be one, leading to

$$\sum_{m=1}^{n_p} v_{i,j}^m \geq 1,\ \forall\ i=1,2,...,n_r,\ j=1,2,...,n_c. \quad (2)$$

*2) Excluding disjoint flow loops*

With the constraints (1) and (2), disjoint loops may appear on the flow paths. For example, the constraint does not prevent the disjoint loop at the lower right side of the valve array in Fig. 6(c) from happening. All the valves and cells on the loop meet the constraints (1) and (2), but this loop gives a false counting of valve coverage in testing, because pressure from the source cannot reach a valve on this loop so that it is not possible to test whether the valves on the loop can be opened.

To solve the disjoint loop problem, we force the air pressure from the source to reach any segment of the path. To represent the pressure volume (pressure flow) passing through a valve at the location $(i,j)$, we define an integer variable $f_{i,j}^m$. This variable is positive when viewed from a cell which the pressure flow enters; it is negative when viewed from a cell which the pressure flow leaves. In addition, a pressure can pass through a valve only if the valve is on the test path, under the condition $v_{i,j}^m=1$. Otherwise $f_{i,j}^m$ should be set to 0. This condition constrains $f_{i,j}^m$ as

$$f_{i,j}^m \leq v_{i,j}^m \cdot \mathcal{M} \quad \text{and} \quad f_{i,j}^m \geq -v_{i,j}^m \cdot \mathcal{M} \quad (3)$$

where $\mathcal{M}$ is a large positive constant [10]. Consequently, the pressure volume in the cell at $(i,j)$ when testing the $m$th flow path is equal to the sum of the volumes of the four surrounding valves. This relation can be written as

$$f_{i,j-1}^m+f_{i,j+1}^m+f_{i+1,j}^m+f_{i-1,j}^m=c_{i,j}^m \quad (4)$$

where the cells on the left of, on the right of, above and below the cell at the location $(i,j)$ are indexed by $(i,j-1)$, $(i,j+1)$, $(i+1,j)$ and $(i-1,j)$, respectively.

Constraint (4) prevents disjoint loops from appearing effectively. Assume there is a disjoint loop on the $m$th path and the cells on the disjoint loop are indexed by $(i_1,j_1)$ to $(i_l,j_l)$, where the valves at $(i_l,j_l)$ and $(i_1,j_1)$ are neighbors. For each cell on the loop we can write a constraint similar to (4). Adding the left and right sides of these constraints together, we have

$$\sum_{(i,j)\in I_l}(f_{i,j-1}^m+f_{i,j+1}^m+f_{i+1,j}^m+f_{i-1,j}^m)=\sum_{(i,j)\in I_l} c_{i,j}^m \quad (5)$$

where $I_l$ is the index set $\{(i_1,j_1)...(i_l,j_l)\}$ for the cells on the loop. On a disjoint loop, the sum on the left side of (5) is always equal to 0, because no pressure flow enters the loop. This contradicts the fact that $\sum_{(i,j)\in I_l} c_{i,j}^m$ should be larger than 0, because the cells are on the flow path. The concept of this model is illustrated in Fig. 6(d).

*3) Finding the minimum set of flow paths*

The path constraints defined above rely on a known number $n_p$ paths that can guarantee the coverage of all valves. In our formulation, we first assign $n_p$ a constant and then try to find a set of paths whose number is no larger than $n_p$ that can cover all valves. For each path, we assign a 0-1 variable $p_m, 1 \leq m \leq n_p$ to indicate whether this path is used. Because any valve on the $m$th path marks the path to be used, $p_m$ can be constrained as

$$p_m \cdot \mathcal{M} \geq \sum_{(i,j)\in I} v_{i,j}^m \quad (6)$$

where $\mathcal{M}$ is a positive constant larger than the number of valves on the array, and $I$ is the index set of all valves. If a valve is on the $m$th path, the right side of (6) is larger than 0, so that $p_m$ must be set to 1 to meet the constraint.

With the constraints above, the ILP problem to find a minimum set of paths that cover all valves can be formulated as follows,

$$\text{minimize} \quad \sum_{m=1}^{n_p} p_m \quad (7)$$

$$\text{subject to} \quad (1)-(4)\ \text{and}\ (6). \quad (8)$$

Since we specify the number of paths $n_p$ as a constant, it is possible that the ILP problem above has no solution, meaning that not all the valves can be covered by $n_p$ flow paths. If this happens, we increase $n_p$ and solve the optimization problem again.

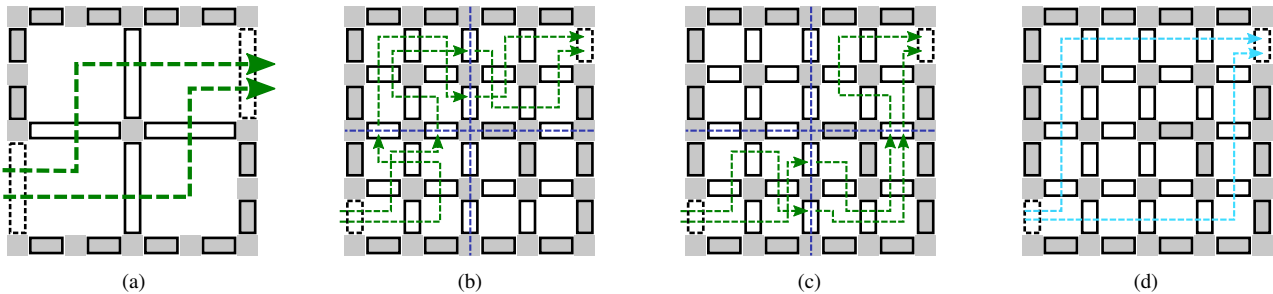

Fig. 7. Hierarchical path construction and valve search for cut-sets. (a) Paths at the top level indicate flow direction. (b)/(c) Subpaths in subblocks forming final flow paths. (d) Searching beginning and ending valves of cut-sets. Vales along the two search directions form the starting and ending valve sets.

*4) Improving scalability with a hierarchical model*

To improve the scalability of the proposed method, we apply a hierarchical approach, where we partition the valve array into subblocks and find the test paths in each subblock individually. Afterwards, the flows paths at the top level are formed by connecting the subpaths at lower levels. With this technique, the proposed method can process large designs more efficiently but generates more test vectors.

The concept of this hierarchical model can be explained using the example in Fig. 7(a)-(c), where the original FPVA is a $4\times4$ array, and we consider each $2\times2$ array as a subblock. Consequently, the top level becomes a $2\times2$ array. Thereafter, we apply the ILP formulation described above to find the top level flow paths, as shown in Fig. 7(a). The flow paths at the top level define the direction of the subpaths in the subblocks. For example, if there is a path from the top left to the top right and the right boundary of the subblock has two valves, there should be at least two paths in the flow direction in the subblocks to cover those two valves. The test paths in each subblock are generated by solving the ILP problem similar to (7)–(8), while constraining further that these subpaths start and end at the boundary of the subblock at the side where the top-level paths enter and leave the subblock. Finally, the final test paths are created by connecting the subpaths in different subblocks, as shown in Fig. 7(b)-(c). The rule of this connection is that a subpath should be included at least once, so that the valves it covers are covered by the final test paths.

*C. Generating cut-set test vectors*

Besides flow paths, we also need to create cut-sets to test whether all valves can be closed. Because a cut-set separates the source and the sink, an end of a cut-set must touch an edge of the chip, as shown in Fig. 5(b) and 5(d). Observing this phenomenon, we search the valves along the boundary of the chip in two directions starting from the source until the sink port is reached from both directions, as shown in Fig. 7(d). Consequently, we find two sets of valves, and a cut-set must include a valve from each of these sets.

The cut-set generation problem is a complementary problem of finding a set of flow paths covering all valves described in Section III-B and can be solved by adapting the optimization problem (7)–(8) to include the additional constraint that a cut-set must start and end at the boundary of the chip.

As discussed in Section III-A, we must prevent the pattern shown in Fig. 5(c) and 5(d) from appearing in the cut-set to guarantee the detection of two faults masking each other. This illegal pattern can be described as a new cut-set can be formed by only one new valve with some valves from the old cut-set. Assume there is a valve at the location $(i,j)$ and the two obstacle areas at the two ends of the valve are indexed by $(i_1,j_1)$ and $(i_2,j_2)$. To prevent this pattern from being formed, we add an additional constraint to the optimization problem as

$$c_{i_1,j_1}^m + c_{i_2,j_2}^m - 1 \leq v_{i,j}^m \tag{9}$$

which specifies that if the two ends of a valve are in the current cut-set, this valve must be included in the cut-set to prevent the illegal pattern in Fig. 5(c) and 5(d).

## IV. SIMULATION RESULTS

The proposed framework was implemented in C++ and tested using a 3.20 GHz CPU with 8 GB memory. We demonstrate the results using FPVAs of different rows and columns. The tested arrays are shown in Table I, where the first column shows the dimensions of the arrays and the second column shows the number of valves. These arrays contain long channels for transportation and obstacle areas without valves.

In the simulation, the dimension of subblocks was set to $5\times5$ and the arrays were partitioned regularly at the top level. The hierarchy of the test cases is shown in the third and the fourth columns in Table I. The numbers of test vectors generated for flow paths and cut-sets are shown in the columns $n_p$ and $n_c$, respectively. Although not explained in detail in this paper, the proposed method can also be adapted to generate test vectors to detect control layer leakage. The numbers of these test vectors are shown in the column $n_l$ in Table I. The total number of test vectors are shown in the column $N$. These numbers are roughly two times of the square root of the numbers of valves in the arrays. Since this is the first method proposed for FPVA fault test, we do not have a way to compare the efficiency of the proposed method. However, consider a simple baseline method where only one valve is switched open or closed each time for fault test. The total number of test vectors in this case would be two times of the number of valves, a squared complexity compared with the proposed method.

The runtimes for generating each set of test vectors are shown in the columns $t_p$, $t_c$, and $t_l$, respectively. The total runtime for generating all these vectors is shown in the column $T$. For generating the flow-path and cut-set vectors, the proposed model was solved in a few minutes. For generating the vectors to test control layer leakage, the largest array used about 25 minutes. The large runtime results from the facts that the problem to find a minimal set of paths in an undirected graph to cover all nodes is NP-hard, and to guarantee the detection of any two faults by excluding the patterns shown in Fig. 5(c) increases the complexity of the problem tremendously. In practice, the proposed method needs to be executed offline only once to generate the test vectors for a given array architecture, so that the runtime is already acceptable. However, the improvement of the efficiency of the model is still our focus, and we are currently testing a vector-based path generation model which can eliminate many variables so that even much larger valve arrays can be processed in a reasonable time.

TABLE I
RESULTS OF TEST VECTOR GENERATION

| Valve Array | | Hierarchy | | Flow Paths | | Cut-sets | | Control Leakage | | Total | |
|---|---|---|---|---|---|---|---|---|---|---|---|
| Dimension | $n_v$ | Top | Subblock | $n_p$ | $t_p(s)$ | $n_c$ | $t_c(s)$ | $n_l$ | $t_l(s)$ | $N$ | $T(s)$ |
| $5\times 5$ | 39 | $1\times 1$ | $5\times 5$ | 5 | 0.3 | 8 | 0.2 | 4 | 2 | 17 | 2.5 |
| $10\times 10$ | 176 | $2\times 2$ | $5\times 5$ | 4 | 4 | 18 | 5 | 4 | 10 | 26 | 19 |
| $15\times 15$ | 411 | $3\times 3$ | $5\times 5$ | 8 | 17 | 28 | 26 | 8 | 127 | 44 | 170 |
| $20\times 20$ | 744 | $4\times 4$ | $5\times 5$ | 16 | 35 | 38 | 41 | 16 | 742 | 70 | 818 |
| $30\times 30$ | 1704 | $6\times 6$ | $5\times 5$ | 20 | 255 | 58 | 171 | 20 | 1492 | 98 | 1918 |

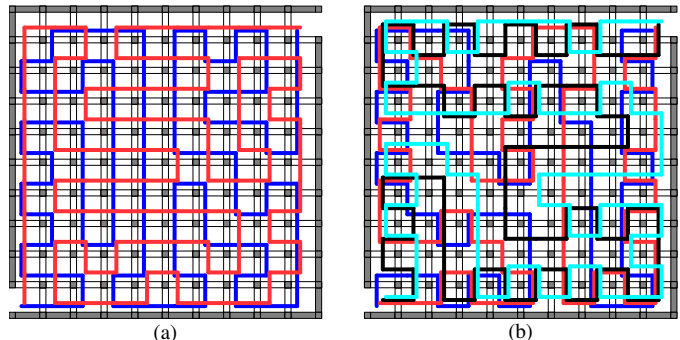

Fig. 8. Comparison of the hierarchical model to the direct model for a $10\times 10$ FPVA. (a) Two flow paths from the direct ILP model. (b) Four paths from the hierarchical model.

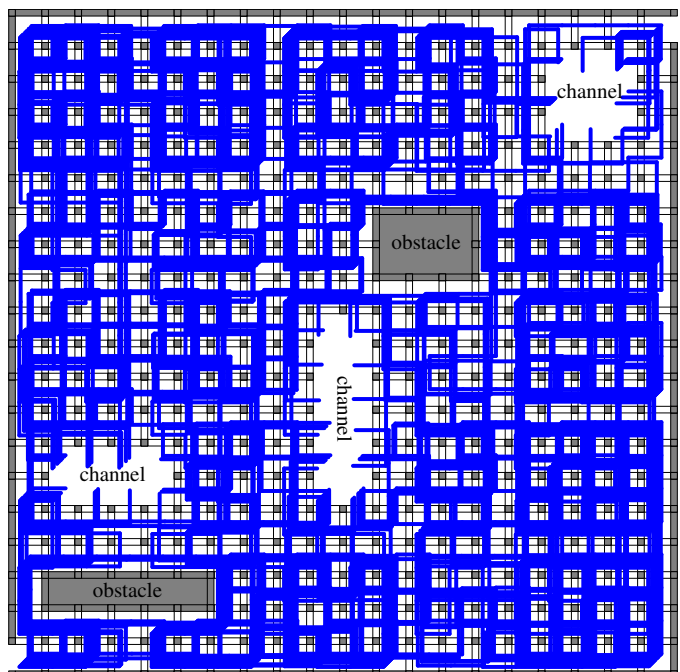

Fig. 9. A total of 16 flow paths for the $20\times 20$ array with channels and obstacles.

To demonstrate the effectiveness of the hierarchical model, we show the flow paths of a $10\times 10$ valve array without channels or obstacles. In Fig. 8(a), the flow paths are generated using the ILP model without hierarchy directly. It is interesting to observe that only two flow paths are needed to cover all the valves in this regular array. In Fig. 8(b) the flow paths for the same array are generated using a hierarchy with the subblock dimension $5\times 5$. The number of paths in this case is four, a little larger than the number from the direct model, but still acceptable.

To demonstrate the generated test vectors, we show the flow paths for the $20\times 20$ valve array from Table I in Fig. 9. With only 16 paths, all the 744 valves in this array are covered. In this array, there are three channels and two obstacles, demonstrating that the proposed method can also deal with FPVAs with irregular structures efficiently.

To verify whether the combination of flow paths and cut-sets can detect faults as explained in Section III, for each valve array in Table I we randomly introduced one, two, three, four and five faults, respectively, and applied the generated test vectors. We repeated this process 10 000 times. In these test cases, the test vectors captured all the faults. Therefore, we can conclude that these test vectors are very effective in practice in detecting faults, although in theory they can only guarantee the detection of two faults in a chip.

## V. CONCLUSION

We have proposed the first strategy to detect manufacturing faults in fully programmable valve arrays (FPVAs). We have also introduced a hierarchical ILP-based model to identify a small set of test vectors. The proposed method can guarantee the detection of any two faults in a chip, and it also demonstrated its potential in detecting more than two faults in a chip in our simulation.

## ACKNOWLEDGMENT

The work of B. Li and U. Schlichtmann was supported by the IGSSE Project FLUIDA of Technical University of Munich. The work of B. B. Bhattacharya was supported, in part, by the special research grant funded by PPEC, Indian Statistical Institute, Kolkata. The work of Chunfeng Liu was supported fully, and the work of K. Chakrabarty and T.-Y. Ho was supported in part, by the Technical University of Munich – Institute for Advanced Study, funded by the German Excellence Initiative and the European Union Seventh Framework Programme under grant agreement n° 291763. The work of T.-Y. Ho was also supported in part by the Ministry of Science and Technology of Taiwan, under Grant MOST 105-2221-E-007-118-MY3.